\documentclass[10pt]{article}
\usepackage{amsmath}
\usepackage{amssymb}
\usepackage{graphicx}
\usepackage[resetlabels]{multibib}
\newcites{main}{References}
\newcites{si}{Supplementary References}
\usepackage{color}

\usepackage[labelfont=bf,labelsep=period,justification=raggedright]{caption}

\makeatletter
\renewcommand{\@biblabel}[1]{\quad#1.}
\makeatother

\date{}

\begin{document}
\begin{flushleft}
{\Large
\textbf{Evolution of opinions on social networks in the presence of competing committed groups}
}
\\
Jierui Xie$^{1,3}$,
Jeffrey Emenheiser$^{2,3}$,
Matthew Kirby$^{2,3}$
Sameet Sreenivasan$^{1,2,3,\ast}$,
Boleslaw K. Szymanski$^{1,3}$,
Gyorgy Korniss$^{2,3}$
\\

\bf{1} Department of Computer Science, Rensselaer Polytechnic Institute, 110 8th Street, Troy NY 12180 USA
\\
\bf{2} Department of Physics, Applied Physics, and Astronomy, Rensselaer Polytechnic Institute, 110 8th Street, Troy NY 12180 USA
\\
\bf{3} Social and Cognitive Networks Academic Research Center, Rensselaer Polytechnic Institute, 110 8th Street, Troy NY 12180 USA
\\
$\ast$ E-mail: sreens@rpi.edu
\end{flushleft}

\section*{Abstract}
Public opinion is often affected by the presence of committed groups
of individuals dedicated to competing points of view. Using a model
of pairwise social influence, we study how the presence of such
groups within social networks affects the outcome and the speed of
evolution of the overall opinion on the network. Earlier work
indicated that a single committed group within a
dense social network can cause the entire network to quickly adopt
the group's opinion (in times scaling logarithmically with the
network size), so long as the committed group constitutes
more than about $10\%$ of the population (with the findings being
qualitatively similar for sparse networks as well). Here we study
the more general case of opinion evolution when two groups committed
to distinct, competing opinions $A$ and $B$, and constituting fractions
$p_A$ and $p_B$ of the total population respectively, are present in the
network. We show for stylized social networks (including
Erd\H{o}s-R\'enyi random graphs and Barab\'asi-Albert scale-free
networks) that the phase diagram of this system in parameter space
$(p_A,p_B)$ consists of two regions, one where two stable
steady-states coexist, and the remaining where only a single stable
steady-state exists. These two regions are separated by two
fold-bifurcation (spinodal) lines which meet tangentially and
terminate at a cusp (critical point). We provide further insights to
the phase diagram and to the nature of the underlying phase
transitions by investigating the model on infinite (mean-field
limit), finite complete graphs and finite sparse networks. For the latter case, we also
derive the scaling exponent associated with the exponential growth
of switching times as a function of the distance from the critical
point.


\section*{Introduction}
Since the seminal work of Gabriel Tarde \citemain{Tarde69} in the late
1800s, the shaping of public opinion through interpersonal influence
and conformity has been a subject of significant interest in
sociology. This topic is especially relevant today due to the
preponderance of online social media where individuals can influence
and be influenced by their numerous and geographically scattered
contacts. Public opinion on an issue is often shaped by the actions
of groups that rigidly advocate competing points of view. The most
evident example of such a process occurs during elections when
multiple parties campaign to influence and win over the majority of 
voters. In this as well as other common scenarios, the
predominant means of influencing public opinion involves some form
of broadcast outreach such as television advertising, public
demonstrations etc. 
However, even though factors exogenous to the network may have a significant effect on individuals becoming informed and engaged in particular issues \citemain{Lehmann2011}, there is reason to believe that large scale changes in behavior or opinion are driven primarily through interpersonal influence events occurring within the network. Specifically in the context of rural campaigns, there is evidence
that interpersonal channels constitute the dominant pathways for
effecting individual behavior change, even when direct external
influence is present \citemain{Morris2000}.
Furthermore, with data on social networks becoming
increasingly accessible, there has been a surge of interest in
understanding how campaigns can be successfully won by leveraging
pathways of social influence within the network, thus diminishing
the need for, or complementing the effect of broadcast outreach.

Motivated by these observations, we study a simple model that
enables us to draw useful insights on the evolution of opinions on a
social network in the presence of two groups within the network that
are committed to distinct, competing opinions on an issue. Within
the limits of our model, one of the questions our work answers is
the following. Suppose the majority of individuals on a social
network subscribe to a particular opinion on a given issue, and
additionally some fraction of this majority are
unshakeable in their commitment to the opinion. Then, what should be
the minimal fractional size of a competing committed group in order
to effect a fast reversal in the majority opinion? In addition to answering this question quantitatively, we
show the existence of two distinct types of phase transitions that
can occur in the space of committed fraction pair values.

 We model the dynamics of social influence using a two-opinion variant of the Naming Game \citemain{Steels1995,Baronchelli2006,DallAsta_PRE2006} which also corresponds to a special case of the game introduced and studied in \citemain{Baronchelli_PRE2007, Castello2009}.  The same model was referred to as the binary-agreement model in \citemain{Xie2011}. In this model, at any time, a
node possesses either one of the two competing opinions (i.e. the
node is in state $A$ or state $B$), or both opinions simultaneously
(state $AB$). In a given time step, we choose a node randomly,
designate it as the {\it speaker} and choose one of its neighbors
randomly and designate it as the {\it listener}. The speaker
proceeds to convey its opinion to the listener (chosen randomly if
it possesses two) to the listener. If the listener possesses this
opinion already, both speaker and listener retain it while
eliminating all other opinions; otherwise, the listener adds the
opinion to his list. A table of possible interactions and outcomes
between node-pairs is provided in Table S1. We emphasize
that each node interacts and is influenced only by its neighbors on
the network. There is no element in our model that represents an
external influence mechanism such as the use of media, public
demonstrations, or door-to-door campaigns by members of the
competing groups. Except for their being un-influencable, the
committed nodes are assumed to be identical in all other respects to
uncommitted nodes. In particular, committed nodes do not influence
their neighbors at a different rate or with a higher strength than
uncommitted nodes.

Opinion dynamics models involving committed individuals all
subscribing to a unique opinion have been studied previously in
\citemain{Xie2011,Mobilia2003,Galam2007,Lu2009}. The situation pertinent
to this paper - that of two competing committed groups - has
received considerably greater attention
\citemain{Galam2007,Galam2010,Mobilia2007, Biswas2009, Yildiz2011}.
Mobilia et al. \citemain{Mobilia2007} studied how the presence of
zealots (equivalent to committed individuals) affected the eventual
distribution of opinions (stationary magnetization) in the case of
the voter model. They demonstrated that the distribution for a
finite sized network was Gaussian, with a width inversely
proportional to the square root of the number of zealots, and
centered at $\frac{z_+ - z_-}{z_+ + z_-}$ where $z_+$,$z_-$,
represent the fraction of zealots in the two competing states.
Similarly to \citemain{Mobilia2007}, Yildiz et al. \citemain{Yildiz2011} studied the
properties of steady-state opinion distribution for the voter model
with {\it stubborn} agents, but additionally considered the optimal
placement of stubborn agents so as to maximally affect the
steady-state opinion on the network. Interestingly, unlike in the
model studied here, in the voter model, no transitions in
steady-state magnetization are observed as the committed fraction
pair values are smoothly varied. Biswas et al. \citemain{Biswas2009} considered the effect of having rigid individuals in a one-dimensional system of binary opinion evolution,
and demonstrated a power-law dependence for the decay of steady-state magnetization on the fraction of rigid individuals.The work done in
\citemain{Galam2007,Galam2010} is similar in spirit to our work here;
however, an important difference is that these studies only
considered the infinite-network size limit for complete graphs. We
study finite networks, both complete and sparse, and provide
semi-analytical arguments regarding timescales that become relevant
when the network size is finite.

\section*{Analysis}
First, we study the mean-field version of the model, also being
equivalent to the dynamics on the complete graph in the limit of
infinite system size. We designate the densities of uncommitted
agents in the states $A$, $B$ and $AB$ by $n_A$, $n_B$ and $n_{AB}$.
We also designate the fraction of nodes committed to state $A$, $B$
by $p_A$, $p_B$ respectively. These quantities naturally obey the
condition: $n_A + n_B + n_{AB} + p_A + p_B = 1$. In the asymptotic
limit of network size, and neglecting fluctuations and correlations,
the system can be described by the following mean-field equations,
for given values of the parameters $p_A$ and $p_B$:
\begin{eqnarray}
\frac{dn_A}{dt} &=& -n_A n_B + n_{AB}^2 + n_A n_{AB} + \frac{3}{2} p_A n_{AB} - p_B n_A \nonumber \\
\frac{dn_B}{dt} &=& -n_A n_B + n_{AB}^2 + n_B n_{AB} + \frac{3}{2} p_B n_{AB} - p_A n_B \nonumber \\
\label{MF}
\end{eqnarray}
The evolution of $n_{AB}$ follows from the constraint on densities
defined above. In general, the evolution of the system depends on
the relative values of $p_A$ and $p_B$. In the case of $p_A > 0$,
$p_B = 0$ (or equivalently, $p_B > 0$, $p_A = 0$) there is only a
single group of committed nodes in the network, all of whom
subscribe to the same opinion. This was the case studied in
\citemain{Galam2007,Xie2011,Lu2009}. In this scenario, a transition is
observed when this committed group constitutes a critical fraction of the total network. Specifically, the transition point
separates two dynamical scenarios in the phase space, $(n_A,n_B)$,
of uncommitted node densities. Below the critical value,  the
absorbing state (e.g., $n_{A}=1-p_{A}$, $n_{B}=n_{AB}=0$ when
$p_A>0,p_B=0$) coexists in phase space with a stable mixed
steady-state and an unstable fixed (``saddle") point. At or above the
critical value, the latter non-absorbing steady-state and 
the saddle point cease to exist. Consequently, for a finite system,
reaching the (all $A$) consensus state requires an exponentially
long time when $p$ is less than the critical value. Beyond the
critical value this time grows only logarithmically with network
size. Note that this critical value or threshold is analogous to a
spinodal point \citemain{Landau2000,Rikvold_PRE2010} associated with an
underlying first-order (or discontinuous) transition in equilibrium systems.

\begin{figure}[!ht]
\begin{center}
\includegraphics[width=4.5in]{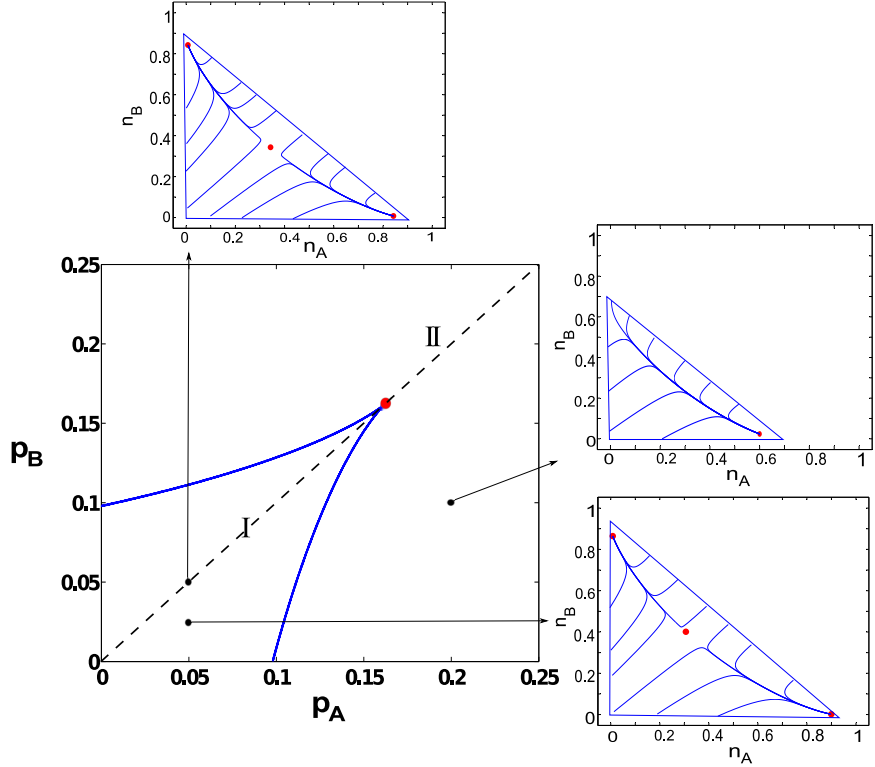}
\end{center}
\caption{ {\bf Mean-field picture in parameter space.} The phase
diagram obtained by integrating the mean-field Eqs.~(\ref{MF}). The
two lines indicate saddle-node bifurcation lines which form the
boundary between two regions with markedly different behavior in
phase space. For any values of parameters within the beak,
denoted as region {\rm I}, the system has two stable fixed points separated by a
saddle point. Outside of the beak, in region {\rm II}, the system has a single stable
fixed point. The saddle-node bifurcation lines meet tangentially and
terminate at a cusp bifurcation point.}
\label{Fig1}
\end{figure}

In order to effectively characterize the behavior of the system
governed by Eqs.~(\ref{MF}) for $p_A,p_B>0$, we systematically
explore the parameter space $(p_A,p_B)$ by dividing it into a grid
with a resolution of $0.000125$ along each dimension. We then
numerically integrate Eqs.~(\ref{MF}) for each $(p_A,p_B)$ pair on
this grid, assuming two distinct initial conditions, $n_A =
1-p_A-p_B, n_B = n_{AB} = 0$ and $n_B = 1-p_A-p_B, n_A = n_{AB} =
0$, representing diagonally opposite extremes in phase space. The
results of this procedure reveal the picture shown in
Fig.~\ref{Fig1} in different regions of parameter space.  As is
obvious, with non-zero values for both $p_A,p_B$, consensus on a
single opinion can never be reached, and therefore all fixed points
(steady-states) are non-absorbing. With $(p_A,p_B)$ values within
the region denoted as ${\rm I}$ which we refer to as the ``beak"
(borrowing terminology used in \citemain{Dykman1994}), the phase space
contains two stable fixed points, separated by a saddle point, while
outside the beak, in region ${\rm II}$, only a single stable fixed point exists in phase
space. In region {\rm I}, one fixed point
corresponds to a state where opinion $A$ is the majority opinion
($A$-dominant) while the other fixed point corresponds to a state where opinion
$B$ constitutes the majority opinion ($B$-dominant). Figure~\ref{Fig1} shows representative trajectories and fixed points in
phase space, in different regions of parameter space. Similar phase
diagrams have been found in other two-parameter systems in different
contexts including chemical reactions \citemain{Dykman1994} and genetic
switches \citemain{Gardner2000}.

In order to study the nature of the transitions that occur when we
cross the boundaries of the beak, we parametrize the system by
denoting $p_B = c p_A$ where $c$ is a real number. Then, we systematically analyze the
transitions occurring in two cases: (i) $c = 1$ and (ii) $c \neq 1$.
It can be shown that along the diagonal line $c=1$ the system
undergoes a cusp bifurcation at $p_A = p_B  = 0.1623$. The movement of the fixed points as $p_A$ and $p_B$ are smoothly varied along the diagonal line is shown in Figure S1. Henceforth,
we denote the value of $p_A$ and $p_B$ at the cusp as $p_c$. As is
well known, at the cusp bifurcation two branches of a saddle-node
(or fold) bifurcation meet tangentially \citemain{Arnold}. These two
bifurcation curves form the boundary of the beak shown in
Fig.~\ref{Fig1}. A detailed analysis demonstrating that $p_A = p_B =
p_c$ constitutes a cusp bifurcation, as well as a semi- analytical
derivation of the bifurcation curves is provided in the Supporting Text S1 (Sections:
1, 2, 3). The cusp bifurcation point is analogous to a second-order (or continuous) critical point
seen in equilibrium systems, while bifurcation curves are
analogous to spinodal transition lines.
\begin{figure}[!htbp]
\begin{center}
\includegraphics[width=5in]{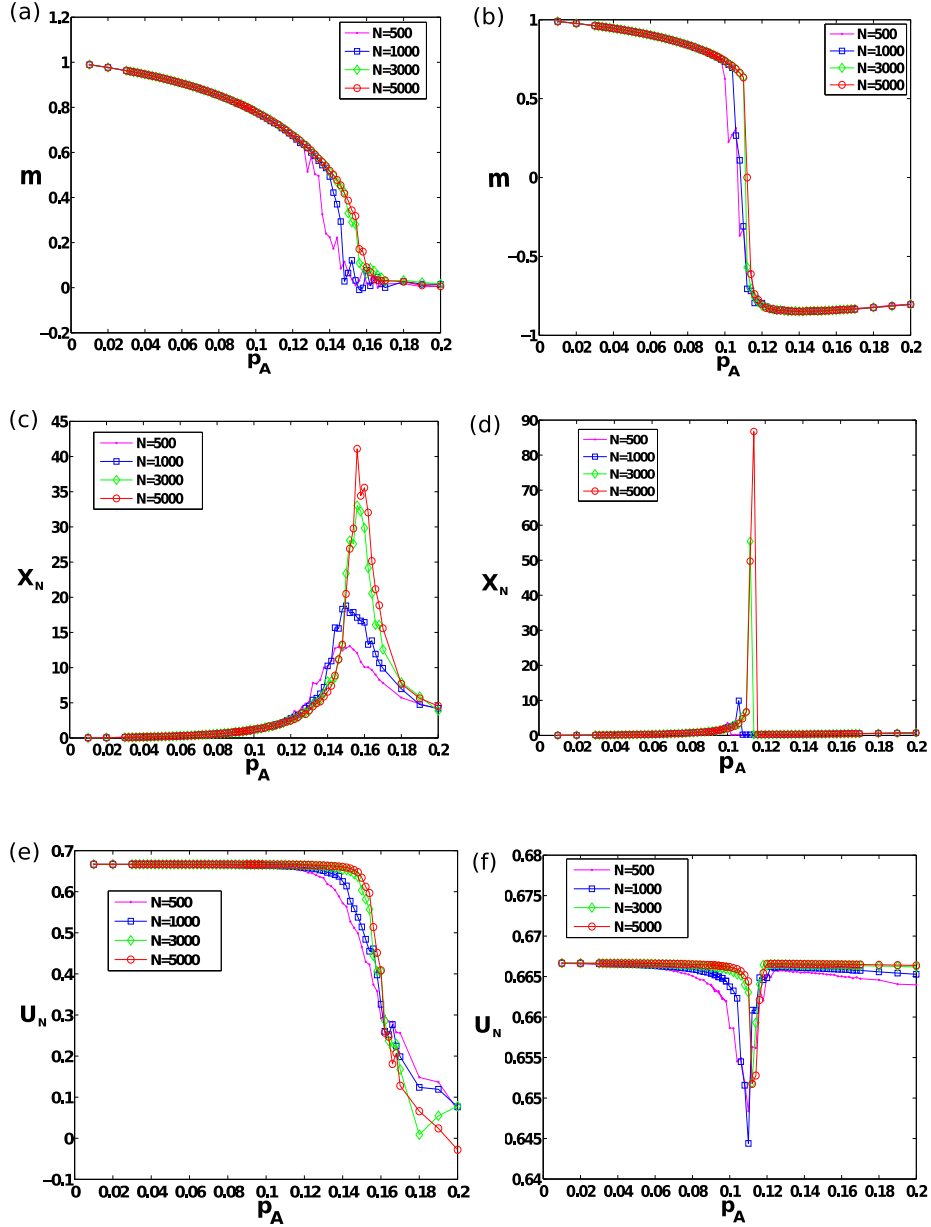}
\end{center}
\caption{
{\bf Behavior of typical order parameters as a function of linear trajectories of slope $c$ that pass through the origin, in parameter space for a complete graph.}  (a)-(b) Steady-state magnetization $m$ defined in the text, for successive $p_A,~p_B$ pairs along lines of slope $c=1$ and $c=0.5$ respectively that pass through the origin. The $c=1$ line in parameter space passes through the cusp point and gives rise to a second-order phase transition, while the $c=0.5$ line passes through a point on the (right) bifurcation line giving rise to a first-order phase transition. Here $10$ realizations of social influence dynamics were performed for each $p_A,~p_B$ pair, starting from the initial condition $n_A = 0,~n_B = 1-p_A-p_B$, and the magnetization was measured conditioned on the system remaining in the steady state that it initially converged to.  (c)-(d) Scaled variance, $X_N$, defined in the text for successive $p_A,p_B$ pairs along lines of slope $c=1$ and $c=0.5$ respectively, that pass through the origin. (e)-(f) Binder cumulant $U_N$ defined in the text for successive $p_A,p_B$ pairs along lines of slope $c=1$ and $c=0.5$ respectively, that pass through the origin. Data for (c),(d),(e) and (f) were generated from $10$ realizations of the social influence dynamics, per $p_A,~p_B$ pair, for each of two initial conditions: $n_A = 1-p_A-p_B,~n_B = 0$ and $n_A = 0,~n_B = 1-p_A-p_B$.
}
\label{Fig2}
\end{figure}

\begin{figure}[!ht]
\begin{center}
\includegraphics[width=4.5in]{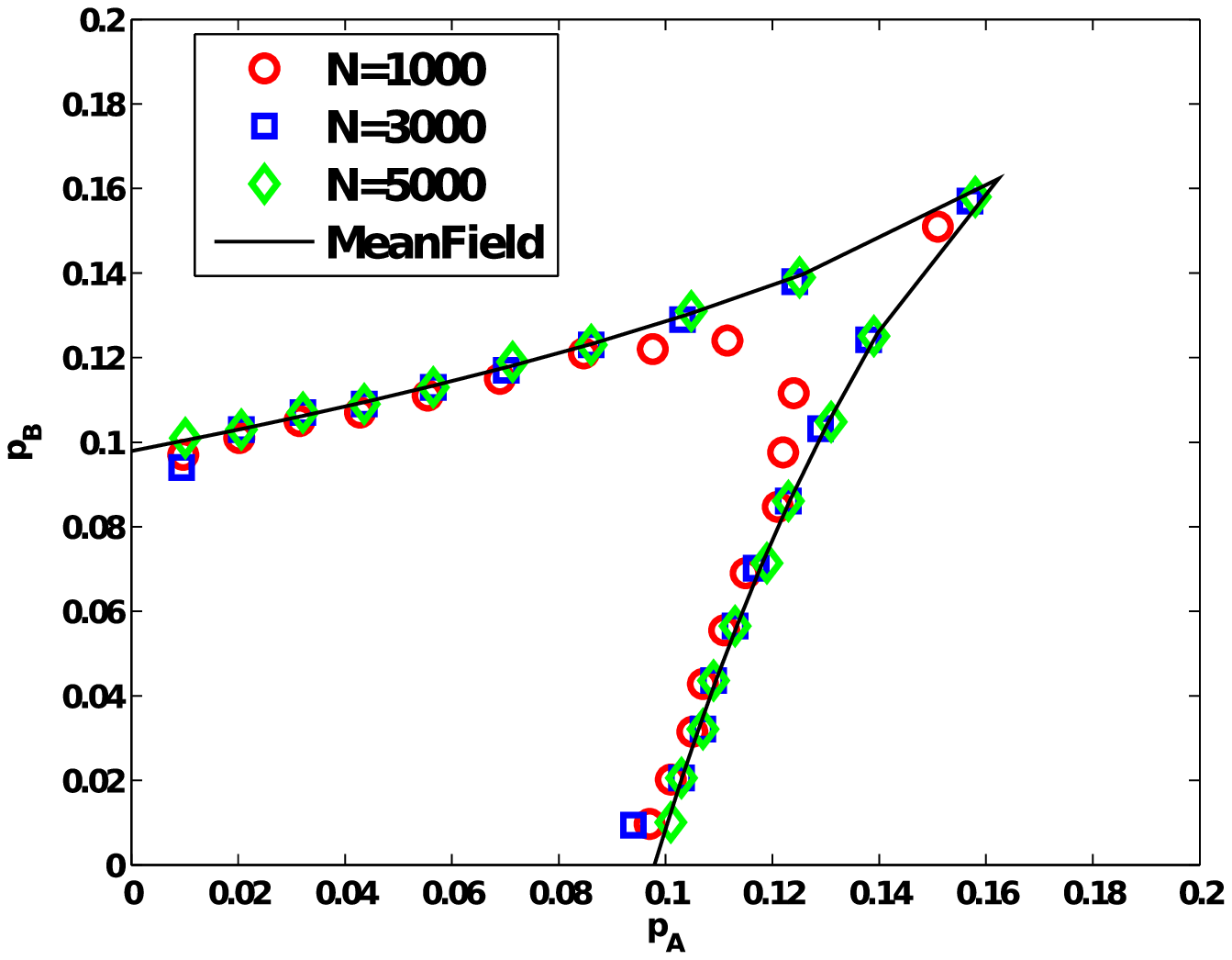}
\end{center}
\caption{
{\bf Picture in parameter space for a complete graph obtained from analytical and simulation results.}
The bifurcation lines and the cusp point in parameter space were obtained analytically from the mean field equations and are compared with those found using simulations for finite-sized complete graphs. Analytical and simulation curves show excellent agreement as $N$ increases. The location of the transition occurring across the bifurcation curve was obtained using the Binder cumulant $U_N$ (Fig. 2(f)), while the location of the cusp point was obtained by using variance of $m$ (Fig. 2(c)). For both analytical and simulation results, the bifurcation curves are obtained by identifying the critical points that lie on linear trajectories in parameter size described by $p_B = c p_A$. This process is carried out for different values of $c$ between $0$ and $1$ at intervals of $0.1$, and for each value of $c$, $p_A$ is varied at a resolution of $0.002$. In simulations, for each such combination of $(p_A,p_B)$ obtained, we perform averages over $10$ realizations of the social influence dynamics, for each of two initial conditions, $n_A = 1-p_A-p_B$, and $n_A = 0$, with $n_B = 1-n_A-p_A-p_B$ for each case.
}
\label{Fig3}
\end{figure}

Next, we study the stochastic evolution of opinions on finite-sized
complete graphs through simulations. Here, we systematically vary
$c$ from $1$ to $0$ to obtain the right bifurcation curve, and
therefore by virtue of the $A\mbox{-}B$ symmetry in the system, also obtain
the left bifurcation curve. In particular for a given value of $c$
we obtain the transition point by varying $p_A$ (with $p_B = c p_A$)
and measuring the quantity:
\begin{equation}
m = (n_B - n_A)/(1-p_A-p_B)
\label{mag}
\end{equation}
which we utilize as an order parameter. The above order parameter is
analogous to the ``magnetization" in a spin system as it captures
the degree of dominance of opinion $B$ over opinion $A$ and is
conventionally used to characterize the nature of phase
transitions exhibited by such a system. 

Another quantity, the Binder cumulant, defined as
\begin{equation}
U_N = 1 - \bigg[ \frac{\langle m^4 \rangle}{3 \langle m^2 \rangle^2} \bigg]
\label{BC}
\end{equation}
for a system of size $N$, is commonly used to distinguish between different types of phase transitions  \citemain{Landau2000}.
The utility of the Binder cumulant comes from the markedly different signatures we expect it to produce
along a spinodal trajectory (e.g. $c = 0.5$) - one that passes through the spinodal line - and one along a trajectory that passes through the critical point (e.g., along the diagonal, $c=1$). This difference arises from the following distinction in the evolution of the distribution of $m$, $P(m)$, along these trajectories. Along a spinodal trajectory starting from a point where $p_A=p_B$, an initially symmetric (about $m = 0$), bimodal $P(m)$ becomes asymmetric and unimodal upon crossing the spinodal line, with the single mode eventually becoming a delta function. In contrast, along the diagonal trajectory in parameter space, $P(m)$ is initially a double-delta distribution (for $p_A = p_B \ll p_c$), symmetric about $m = 0$, and it smoothly transitions to a zero-centred gaussian distribution as the critical point is crossed. The definition of $U_N$ indicates that $U_N = 2/3$ for a delta function distribution (also for a symmetric, double-delta distribution about $m = 0$), while $U_N = 0$ for a zero-centered Gaussian distribution, and thus readily yields the limiting $U_N$ values at both extremes of the spinodal and diagonal trajectory. As illustrated in Fig.~\ref{Fig2}, $U_N$ as a function of $p_A$ shows distinct behaviors for $c = 1$ and $c = 0.5$, indicating the existence of a second-order (or continuous) transition point
at $p_A = p_B = p_c(N)$ (Fig.~\ref{Fig2}(e)) and first-order (or discontinuous) phase transition points (Fig.~\ref{Fig2}(f)) along off-diagonal trajectories \citemain{Landau2000}, respectively. The second-order critical point $p_c(N)$ converges to the mean-field value, $p_c \approx 0.1623$, as $N$ becomes larger. The dip observed in $U_N$ along the off-diagonal trajectory serves as an excellent estimator of the location of the first-order (spinodal) transition for a finite network. Thus, to reiterate, for a finite network, the second-order transition point and
the first-order transition (spinodal) lines are respective analogues of the cusp bifurcation point and the saddle-node bifurcation curves
observed in the mean-field case.

The fluctuations of the quantity $m$ can also be used to identify a transition point, particularly for the case of the second-order transition. In particular, in formal analogy with methods employed in the study of equilibrium spin systems, the scaled variance:
\begin{equation}
X_N = N \langle (|m| - \langle |m| \rangle)^2 \rangle
\label{mvar}
\end{equation}
serves as an excellent  estimate for the second-order transition point $p_c$ for a finite network. As shown in Fig.~\ref{Fig2}(c), $X_N$ peaks at a particular value of $p_A$, with the size of the peak growing with $N$ (and expected to diverge as $N \to \infty$).  In the case of the spinodal transition, one studies fluctuations of $m$ ($X_N = N \langle (m - \langle m \rangle )^2 \rangle$) restricted to the metastable state \citemain{Klein1982,Ray1991} until the spinodal point (Fig.~\ref{Fig2}(d)) at which the metastable state disappears, and fluctuations of $m$ in the unique stable state beyond the spinodal point (Fig.~\ref{Fig2}(d)).

Figure~\ref{Fig3} shows the bifurcation (spinodal) lines obtained via simulations of finite complete graphs by using the Binder cumulant
(Fig.~\ref{Fig2}(f))to identify the location of the spinodal phase transition, and demonstrates that its agreement
with the mean-field curves improves as $N$ grows. The cusp points shown here are identified in simulations as the locations where $X_N$ reaches its
peak value (Fig.~\ref{Fig2}(c)).

\begin{figure}[!ht]
\begin{center}
\includegraphics[width=4.5in]{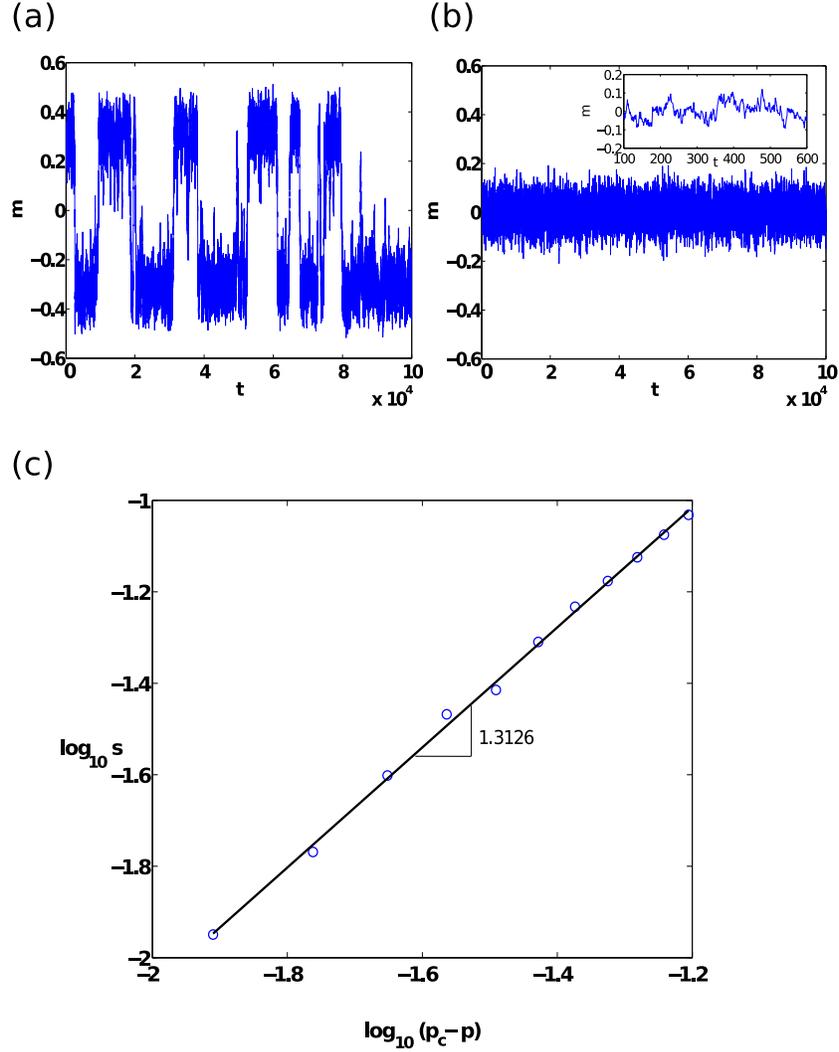}
\end{center}
\caption{
{\bf Evolution of order parameter $m$ and the exponential growth in switching time as a function of distance from the second-order critical point.} (a) Switches in the value of $m$ as a function of time $t$ for a sample evolution (with initial transient removed) of the system when $p_A = p_B = 0.154$ ($<p_c$). This reflects the system repeatedly switching between the $A$-dominant steady-state ($m>0$) and the $B$-dominant steady-state ($m < 0$). (b) Sample evolution of the system (with initial transient removed) for $p_A = p_B = 0.2$ ($>p_c$). The system fluctuates randomly about the only existing steady-state in which densities of $A$ and $B$ nodes are equal. (c) The dependence of $s$ in the exponential scaling $T_{\rm switching} \sim \exp(s N)$ when $p_A = p_B = p$ ($p < p_c$) as a function of $(p_c - p)$, obtained using the eikonal approximation (see SI: Section 4)}
\label{Fig4}
\end{figure}

In the region within the beak, the switching time between the co-existing steady-states represents the longest time-scale of relevance in the system. The switching time is defined as the time the system takes to escape to a distinct co-existing steady-state, after having been trapped in one of the steady-states (see Fig.~\ref{Fig4}(a)). In stochastic systems exhibiting multistability or metastability, it is well known that switching times increase exponentially with $N$ for large $N$ (the weak-noise limit) \citemain{Maier1992, Graham1984, Gang1987, Dykman1994}  Furthermore, the exponential growth rate of the switching time in such cases can be determined using the eikonal approximation \citemain{Dykman1994,Luchinsky1998}. The basic idea in the approximation involves (i) assuming an eikonal form for the probability of occupying a state far from the steady-state and (ii) smoothness of transition probabilities in the master equation of the system. This allows the interpretation of fluctuational trajectories as paths conforming to an auxilliary Hamilton-Jacobi system. This in turn enables us to calculate the probability of escape allowing an {\it optimal fluctuational path} that takes the system from the vicinity of the steady-state to the vicinity of the saddle point of the deterministic system. The switching time is simply the inverse of the probability of escape along this optimal fluctuational path. We defer details of this procedure to Supporting text S1: Section 4. Using this approach we find that for the symmetric case, $p_A = p_B=p < p_c$, the exponential growth rate of the switching time $s \sim (p_c -p)^\nu$ with $\nu \approx 1.3$ (Fig.~\ref{Fig4} (c) ).  Thus, along the portion of the diagonal within the beak:
\begin{equation}
T_{\rm switching} \sim \exp [(p_c - p)^\nu N ]
\label{switchingtime}
\end{equation}
Outside the beak, the time to get arbitrarily close to the sole steady-state value grows logarithmically with $N$ (not shown).

The results presented so far show that there
exists a transition in the time needed by a committed minority to
influence the entire population to adopt its opinion, even in the presence of a committed opposition (i.e. in the
case where both $p_A,p_B >0$), as long as $p_A, p_B < p_c$. (Note that the case $p_A>0,p_B = 0$ was
considered in \citemain{Xie2011}).  For example, assume that initially
all the uncommitted nodes adopt opinion $B$, and that $p_A = p_B <
p_c$. Then, the steady-state that the system reaches in $\ln (N)$ time
is the one in which the majority of nodes hold opinion $B$. Despite
the fact that there exist committed agents in state $A$ continuously
proselytizing their state, it takes an exponentially long time
before a large (spontaneous) fluctuation switches the system to the
$A$-dominant steady-state. For identical initial conditions, the
picture is qualitatively the same if we increase $p_A$ keeping
$p_B$ fixed, as long as $(p_A,p_B)$ lies within the beak. However,
when $(p_A,p_B)$ lies on the bifurcation curve or beyond, the
$B$-dominant steady-state vanishes, and with the same initial
conditions - where $B$ is the initial majority - it takes the system
only $\ln (N)$ time to reach the $A$-dominant state (the only existing
steady-state). Thus, for every value of an existing committed
fraction $p_B$ ($<p_c$) of $B$ nodes, there exists a corresponding
critical fraction of $A$ nodes beyond which it is guaranteed that
the system will reach an $A$ dominant state in $\ln (N)$ time,
irrespective of the initial conditions.
However, for any trajectory in the parameter space in a region where either $p_A$ or $p_B$ is (or both are) greater than $p_c$, no abrupt
changes in dominance or consensus times are observed. Instead, the dominance of $A$ or $B$ at the single fixed point smoothly varies as
the associated committed fractions are varied. Moreover, the system always reaches this single fixed point in $\ln (N)$ time.

\begin{figure}[!ht]
\begin{center}
\includegraphics[width=6in]{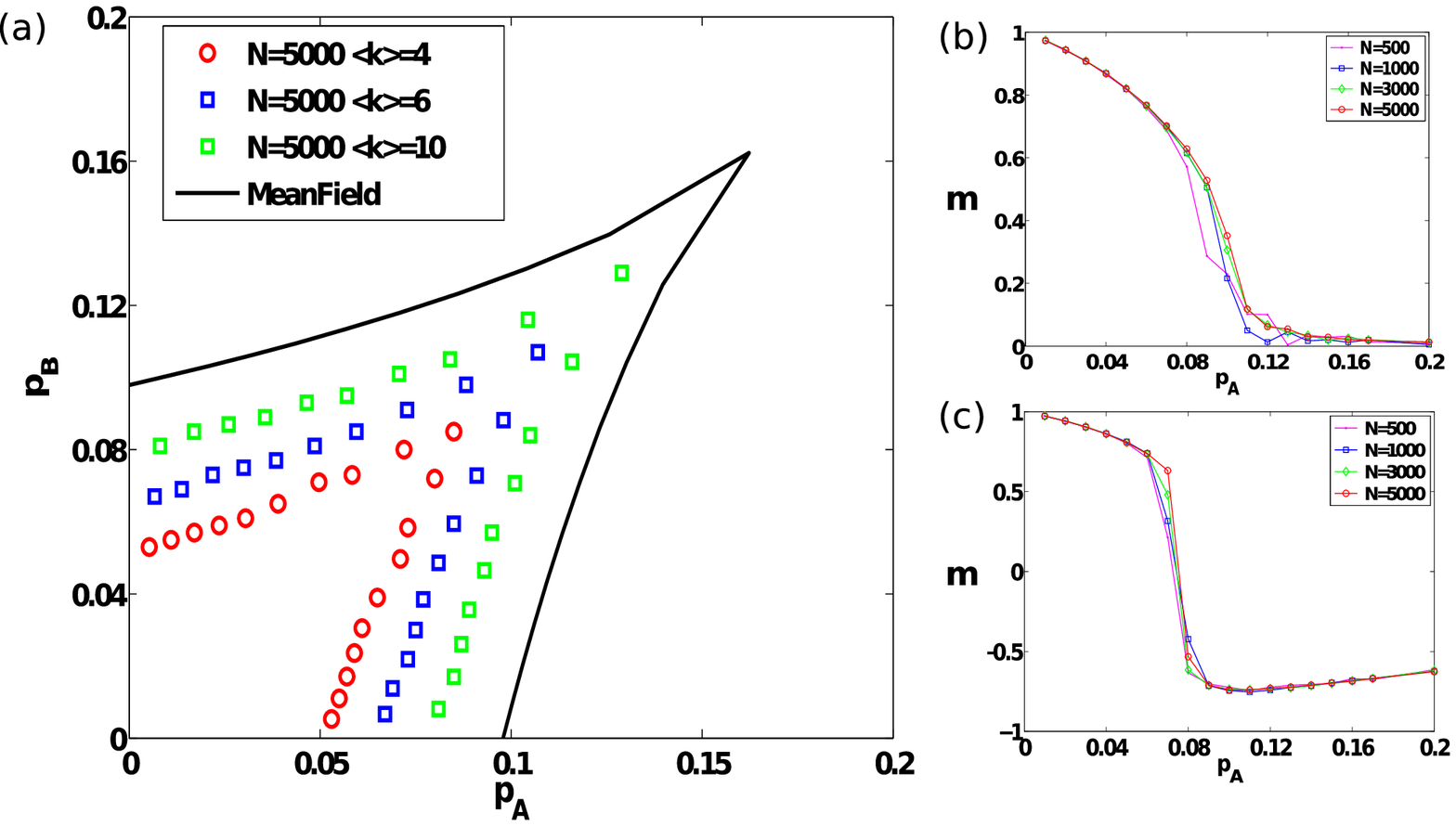}
\end{center}
\caption{
{\bf Results for Erd\H{o}s-R\'enyi random graphs.}(a)The bifurcation lines and cusp point in parameter space obtained through simulations of Erd\H{o}s-R\'enyi random graphs of size $N=5000$ with different average degrees. The mean-field analytical curve is shown for comparison. For simulation results, the bifurcation curves are obtained by identifying the critical points that lie on linear trajectories described by $p_B = c p_A$ in parameter space. This process is carried out for different values of $c$ between $0$ and $1$ at intervals of $0.1$, and for each value of $c$, $p_A$ is varied at a resolution of $0.002$.  For each such combination of $(p_A,p_B)$ obtained, we perform averages for quantities of interest over $10$ realizations of networks (with a single realization of the social influence dynamics per network), for each of two initial conditions, $n_A = 1-p_A-p_B$ and $n_A = 0$ with $n_B = 1-n_A-p_A-p_B$ in each case. (b)-(c) Steady-state magnetization for ER graphs with $\langle k \rangle = 6$ and different sizes $N$ , as parameter pair values are varied successively along slope $c=1$ and  slope $c = 0.5$ lines in parameter space respectively.
}
\label{FigER}
\end{figure}

\begin{figure}[!ht]
\begin{center}
\includegraphics[width=6in]{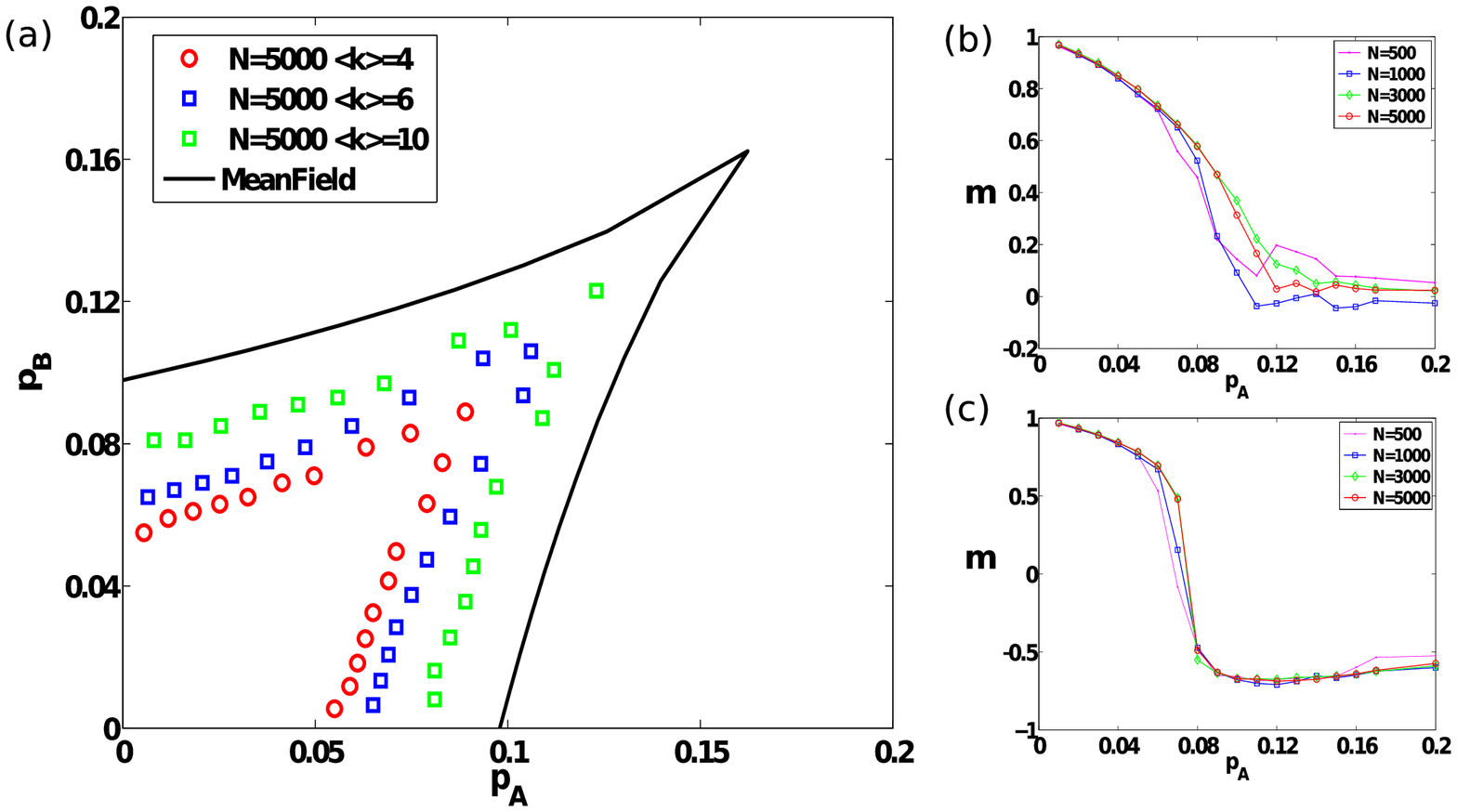}
\end{center}
\caption{
{\bf Results for Barab\'asi-Albert networks.}  (a)The bifurcation lines and cusp point in parameter space obtained through simulations of Barab\'asi-Albert networks of size $N=5000$ with different average degrees. For simulation results, the bifurcation curves are obtained by a similar method as described in the legend of Fig.~\ref{FigER}(a).  (b)-(c) Steady-state magnetization for BA networks with $\langle k \rangle = 6$ and different sizes $N$ , as parameter pair values are varied successively along slope $c=1$ and  slope $c = 0.5$ lines in parameter space respectively. }
\label{FigBA}
\end{figure}

\begin{figure}[!htbp]
\begin{center}
\includegraphics[width=3.75in]{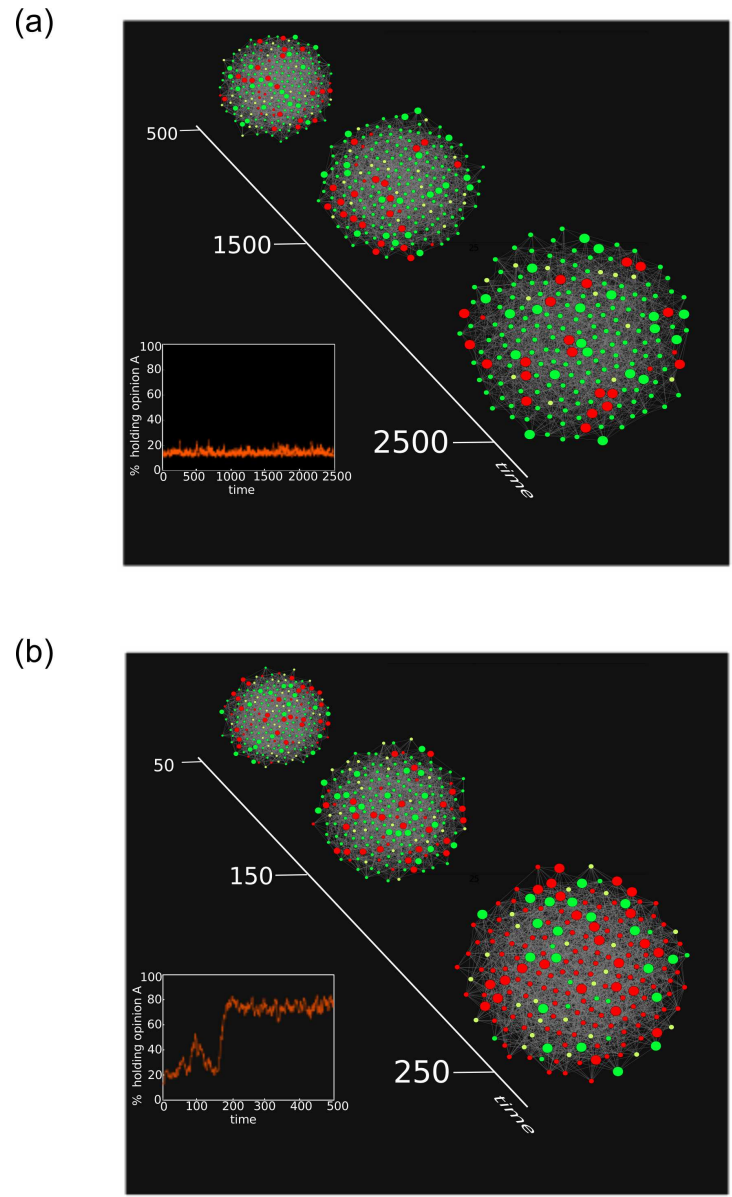}
\end{center}
\caption{
{\bf Visualization of opinion evolutions.}  The evolution of opinions on an ER random graph with $N = 200$ and $\langle k \rangle = 6$ for two $(p_A, p_B)$ pairs. In each case $n_B = 1- p_A - p_B$ and $n_A = 0$. Nodes holding opinion $A$ are depicted in red, while nodes holding opinion $B$ are shown in green. Nodes with larger diameters are committed nodes. Top: The case $p_A = p_B = 0.1$ for which the system is in region {\rm I} in parameter space (following the terminology of Fig. ~\ref{Fig1}, and the system is trapped in a $B$-dominant steady-state. Even after $2500$ time steps, the system continues to remain trapped in this state (inset) with $n_A \approxeq 0.05$.  Bottom: The case $p_A = 0.125$ and $p_B = 0.1$ for which the system is in the region {\rm II}, and undergoes an abrupt transition (inset) to the $A$-dominant state within $250$ time steps.}
\label{Figviz}
\end{figure}

Finally, we study how opinions evolve in the presence of committed
groups on sparse graphs, most relevant to social networks. We study
Erd\H{o}s-R\'enyi (ER) random graphs \citemain{Bollobas} as well as
Barab\'asi-Albert networks \citemain{Barabasi1999}. For each of these
sparse networks, we find the same qualitative behavior as found for
the complete graph. As shown in Figs.~\ref{FigER},~\ref{FigBA}, as
the average degree of the sparse networks increases, the bifurcation
lines in parameter space tend to approach their mean-field
counterparts. 
Although we do not study sparse networks analytically here, we note that in another instance of a phase transition for a similar model studied in \citemain{Baronchelli_PRE2007}, it was demonstrated using heterogeneous mean-field equations that the behavior of sparse networks is qualitatively similar to that of complete graphs. Figure~\ref{Figviz} visually depicts typical instances
of the evolution of opinions on an ER random graph for $(p_A,p_B)$
values within and outside the beak.

\vspace{5mm}
\section*{Discussion}

Using a simple model, we have explored and quantified possible outcomes for the evolution of opinions on a social network in the presence of groups committed to competing opinions. Broadly speaking, our results indicate that as long as the fraction, $p_B$, of nodes committed to a given opinion $B$ is held fixed at a value less than a critical value $p_c$, it is possible to induce the network to quickly tip over to a state where it widely adopts a competing opinion $A$, by introducing a fraction of nodes committed to opinion $A$. The value of the competing committed fraction, $p_A$, at which this tipping point arises depends on the value of $p_B$, and is determined by the bifurcation curve (see Fig.~\ref{Fig1}). Importantly, for a given value of $p_B < p_c$, the excess commitment $p_A-p_B$ required for the network to tip over to $A$ is a decreasing
function of $p_B$ that reaches zero when $p_B = p_c$. While the critical value $p_c$ itself may depend on the network structure and its size, the feature described above holds for the three different classes of networks studied here. A corollary to this feature is that if the committed fraction $p_B$ is held fixed at a value greater than pc, increasing the competing committed fraction $p_A$ only yields continuous incremental gains in the adoption of A (i.e., no tipping point or discontinuous changes in opinions exist).
We analytically determine that $p_c = 0.1623$ for infinite-sized complete graphs, which as observed from our simulation results in Figs. \ref{FigER},~\ref{FigBA} appears to constitute a good upper bound to the value of $p_c$ for sparse networks.

Our results could be of utility in situations where public opinion is deadlocked due to the influence of competing committed groups. Perhaps one example of such a situation is the observed lack of consensus in the U.S. on the existence of human-induced climate change. Indeed, there is evidence in this particular case that the commitment of individuals to particular political ideologies may have an effect on their opinions \citemain{Leiserowitz2011}.

Another scenario to which our model could bear some relevance is the adoption of competing industrial standards. Particularly in situations where a network of entities collaborate or are interdependent, there is a natural attempt at agreement in standards or protocols between interacting members. A classic example of this scenario is the case of the Sellers' screw manufacturing standard that proliferated despite competition from the Whitworth standard \citemain{Sinclair1969}. A key factor responsible for the eventual success of the Sellers standard was William Sellers' leveraging of his connections to corporations and manufacturers \citemain{Wired2002}, whom he persuaded to become adopters of his standard . Furthermore, the network of interdependencies between industries at the dawn of the mass-manufacturing era played an important role in the adoption of the standard becoming widespread. It should be pointed out that in this case, the uncommitted members of the population initially adhered to neither standard - this situation can however be accommodated in our model by assigning each uncommitted his own unique ``opinion" to begin with in close analogy to initial conditions for the original Naming Game \citemain{Steels1995,Baronchelli2006}. 

 A more recent example of such a scenario is the competition between Flash and HTML5 in web-development. There is speculation that Flash, which until recently was the predominant platform for animated web content, is gradually ceding its dominance to HTML5 as a result of the increasing market-share of Apple's mobile devices which exclusively support the latter \citemain{Forbes2011}.
 
 A potential competition between DC fast charging standards is also expected as electric vehicles become increasingly popular with consumers. The front-runners in the mass manufacture of electric vehicles have opted for the CHAdeMO standard, and charging stations compatible with the standard have begun proliferating in the US, Europe and Japan \citemain{SciAm2011,nytimes2011-2}. An alternative to CHAdeMO currently being developed by the Society for Automotive Engineers (SAE), which governs the development of standards in the US automotive industry, is being touted by some car manufacturers as more cost effective as well as technologically superior. However, by the time the first cars employing the SAE standard hit the market, CHAdeMO charging stations are expected to be rather widespread, thus making a competition between the two inevitable \citemain{nytimes2011-3,SciAm2011}. As new collaborations are forged between car-makers especially in the area of electric vehicle development \citemain{wsj2011,nytimes2011} (in addition to collaborations with energy suppliers), the outcome of this competition could be significantly influenced by manufacturers who are already committed to one of the two standards through their investment in them.

To conclude, we have presented results from a simple, abstract model for understanding how opinions on a social network evolve through social influence when there are multiple groups within the network dedicated to competing opinions. Despite the simplicity of our model, we believe the insights provided here form a useful theoretical complement to data-driven studies \citemain{Madan2011} and randomized evaluations \citemain{Banerjee2011} aimed at understanding the spread of opinions.

{\bf Note:} Subsequent to this paper's initial posting on arxiv.org on 12/31/2011 and its acceptance for publication with minor revisions on 1/26/2012, S. Jolad sent us independent unpublished results by D. Linford, P. Hochendoner, A. Reagan, and S. Jolad addressing competing committed groups on the complete graph.

\vspace{5mm}




\newpage

%
%

\title{Supporting Text S1 \\ Evolution of opinions on social networks in the presence of competing committed groups} 
\author{Jierui Xie, Jeffrey Emenheiser, Matthew Kirby, \\ Sameet Sreenivasan, Boleslaw K. Szymanski and Gyorgy Korniss}
\date{}
\maketitle

\renewcommand{\figurename}{Supplementary Figure}
\renewcommand{\tablename}{Supplementary Table}
\setcounter{figure}{0}
\setcounter{equation}{0}

\tableofcontents

\newpage

\section{Analysis of steady states for $p_A = p_B$: existence of a critical value $p_A = p_B = p_c$. }

For notational simplicity we replace $n_A$ by $x$ and $n_B$ by $y$.  The mean field equations describing the system with $p_A = p_B = p$, $0 < p \le 0.5$ then are:
\begin{eqnarray}
\frac{dx}{dt} &=& -x y + (1-x-y-2p)^2 + x (1-x-y-2p) + \frac{3}{2} p (1-x-y-2p) - p x \nonumber \\
\frac{dx}{dt} &=& -x y + (1-x-y-2p)^2 + y(1-x-y-2p) + \frac{3}{2} p (1-x-y-2p) - p y \nonumber \\
\label{SI_MF}
\end{eqnarray}
where $n_{AB} = 1-x-y-2p$. In the steady state, $dx/dt = dy/dt = 0$, and the resulting equations can be solved to yield four solutions for $(x,y)$. Out of these one solution lies outside the valid range for all feasible values of $p$, i.e., $0 < p \le 0.5$. The valid fixed points for Eqs.~\ref{SI_MF} are:

\begin{eqnarray*}
x_1 = \frac{3}{2} - \frac{1}{2} \sqrt{5-2p} - p \nonumber \\
y_1 = \frac{3}{2} - \frac{1}{2} \sqrt{5-2p} - p \nonumber \\
\end{eqnarray*}

\begin{eqnarray*}
x_2 = \frac{1}{2} + \frac{1}{2} \sqrt{1-p^2-6p} - \frac{3}{2}p \nonumber \\
y_2 = \frac{1}{2}  - \frac{1}{2} \sqrt{1-p^2-6p} - \frac{3}{2}p \nonumber \\
\end{eqnarray*}

\begin{eqnarray*}
x_3 = \frac{1}{2} - \frac{1}{2} \sqrt{1-p^2-6p} - \frac{3}{2}p \nonumber \\
y_3 = \frac{1}{2}  + \frac{1}{2} \sqrt{1-p^2-6p} - \frac{3}{2}p \nonumber \\
\end{eqnarray*}

Since the solutions are symmetric in $x$ and $y$, in order to investigate the range of $p$ over which these solutions
are valid, we restrict our analysis to $y$. The solution $y_1$ is valid for all values of $p$. For $y_2,y_3$ to be valid solutions, we require $U(p) = 1-p^2-6p \ge 0$. $U(p)$ is a monotonically
decreasing function for $p > 0$, and the value of $p$ at which $U(p)$ first crosses zero is the critical point.
\begin{equation}
p_c = \sqrt{10}- 3\approxeq 0.1623.
\label{criticalpoint}
\end{equation}
Thus, there exist three fixed points in the range $[0,p_c]$. In the range $(p_c,0.5]$ only one valid fixed point exists, viz. $(x_1,y_1)$.

We can further examine the stability of the obtained fixed points. Linear stability analysis yields the following stability matrix:
\begin{equation}
Q = 
\left[\begin{array}{cc}
-1-\frac{p}{2} & -2+2y^* + \frac{5}{2}p \\
-2+2x^*+\frac{5}{2}p & -1-\frac{p}{2} \\
\end{array}
\right]
\label{stabilitymat}
\end{equation}
where $(x^*,y^*)$ is the fixed point under consideration. 

\begin{table}
\centering
\begin{tabular}{|c|c|} \hline
Before interaction & After interaction \\ \hline
A $\overset{A}{\rightarrow}$   A & A - A \\ \hline
A $\overset{A}{\rightarrow}$   B & A - AB\\ \hline
A  $\overset{A}{\rightarrow}$   AB & A - A\\ \hline
B  $\overset{B}{\rightarrow}$   A & B - AB \\ \hline
B  $\overset{B}{\rightarrow}$   B & B - B\\ \hline
B  $\overset{B}{\rightarrow}$   AB & B - B\\ \hline
AB  $\overset{A}{\rightarrow}$   A & A - A \\ \hline
AB  $\overset{A}{\rightarrow}$   B & AB - AB\\ \hline
AB  $\overset{A}{\rightarrow}$   AB & A - A\\ \hline
AB  $\overset{B}{\rightarrow}$   A & AB - AB \\ \hline
AB  $\overset{B}{\rightarrow}$   B & B - B\\ \hline
AB  $\overset{B}{\rightarrow}$   AB & B - B\\ \hline
\end{tabular}
\label{interactions}
\caption{Shown here are the possible interactions in the binary agreement model. Nodes can possess opinion $A$, $B$ or $AB$, and opinion updates occur through repeated selection of speaker-listener pairs. Shown in the left column are the opinions of the speaker (first) and listener (second) before the interaction, and the opinion voiced by the speaker during the interaction is shown above the arrow. The column on right shows the states of the speaker-listener pair after the interaction.}
\end{table}

The eigenvalues of the stability matrix at the fixed point are given by :
\[
\lambda = -(2+p) \pm \sqrt{26p^2 +(20(x^*+y^*)-36) + 16(1-x^*-y^* +x^*y^*)} ,
\] and examination of the real part of these eigenvalues indicates that $(x_2,y_2)$ and $(x_3,y_3)$ are stable
fixed points, and $(x_1,y_1)$ is an unstable fixed point (saddle point) for $p \le p_c = 0.1623$.
For $p > p_c$, $(x_1,y_1)$, the only valid fixed point, is a stable fixed point.
Supporting Figure 1 shows the movement of the fixed points in the phase space as a function of $p$.

\begin{figure}[htbp]
\begin{center}
\includegraphics[width = 4.5in]{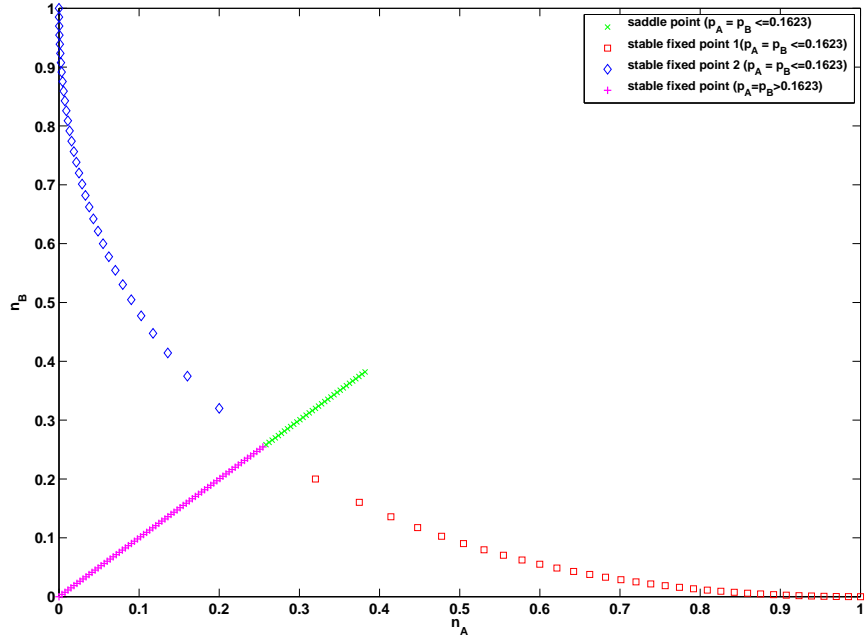}
\end{center}
\caption{{\bf Movement of fixed points as $p_A$ and $p_B$ are smoothly varied along the diagonal line $p_A = p_B$.} For $p_A = p_B , p_c \approxeq 0.1623$ three fixed points exist, two of which are stable, and the third is unstable. For $p_A = p_B > p_c$, only a single stable fixed point exists.}
\label{FigS1}
\end{figure}

\section{Existence of a cusp point}
\label{sec:cusppoint}

Suppose that a \textit{one}-dimensional parameter($\alpha$) dependent system 
\begin{equation}
\label{eq:genSystem}
			\frac{d x}{d t}=f(x;\alpha),   x\in \Re^1, \alpha \in \Re^m
\end{equation}
with smooth function $f$, has an equilibrium at $x=0$ for $\alpha=0$, and let $f_x(0;0)=0$ and $f_{xx}(0;0)=0$ hold. Further, assume that the non-degeneracy conditions (e.g., $f_{xxx}(0;0)\neq 0$) are satisfied. Then the system undergoes a \textit{cusp} bifurcation at $x = 0$ \citesi{Arnold}. 

We prove that such a cusp bifurcation is encountered in our system (i.e., Eq. \ref{SI_MF}) at $p_A = p_B = p_c$ as we move along the diagonal in parameter space ($p_A=p_B$). Note that our system is two-dimensional. To be able to use the above theory, we first need to reduce the dimensionality of our system. The Center Manifold Theorem \citesi{Carr} guarantees the existence of a one-dimensional center manifold to which we can restrict our system, and such a system preserves the same behavior as the original system in the vicinity of the steady-state under consideration. Once we get the restricted system, we can perform the usual bifurcation analysis in one-dimensional system. Following this idea, we first shift the coordinates such that the origin is located at the critical point we found from the $p_B=p_A$ case (for simplicity, we denote $p_A$ by $p$ and $p_B$ by $r$), i.e., $(x_0,y_0;p_0,r_0)=(0.2565,0.2565;\sqrt{10}-3,\sqrt{10}-3)$. 
In the shifted coordinates, the eigenvalues and eigenvectors are given by $\Lambda=[0; -2.1623]$ and
$V =[-0.7071, 0.7071; 0.7071, 0.7071]$. Using the transformation $[\tilde{x} \; \tilde{y}]^{T}=V [x \; y]^{T}$, and after some algebraic manipulations, we obtain in the new co-ordinate system:
\begin{eqnarray}
\label{eq:newSystem2}
\frac{d\tilde{x}}{dt}&=&0.7071(1.5p + 0.2434)(p + r + 1.414\tilde{y} - 0.1623) \nonumber\\
& & - 0.7071(1.5r + 0.2434)(p + r + 1.414\tilde{y} - 0.1623) \nonumber \\
& & - 0.7071(0.7071\tilde{x} + 0.7071\tilde{y} + 0.2566)(p + r + 1.414\tilde{y} - 0.1623) \nonumber \\
& & - 0.7071(p + 0.1623)(0.7071\tilde{x}+ 0.7071\tilde{y} + 0.2566) \nonumber \\
& & + 0.7071(0.7071\tilde{y} - 0.7071\tilde{x} + 0.2566)(p + r + 1.414\tilde{y} - 0.1623)\nonumber \\
& &  + 0.7071(r + 0.1623)(0.7071\tilde{y} - 0.7071\tilde{x} + 0.2566) \nonumber \\
\frac{d\tilde{y}}{d t}&=& - 0.7071(1.5r + 0.2434)(p + r + 1.414\tilde{y} - 0.1623) \nonumber \\
& & - 0.7071(1.5p + 0.2434)(p + r + 1.414\tilde{y} - 0.1623) \nonumber \\
& & - 0.7071(0.7071\tilde{x} + 0.7071\tilde{y} + 0.2566)(p + r + 1.414\tilde{y} - 0.1623) \nonumber \\
& & - 0.7071(p + 0.1623)(0.7071\tilde{x} + 0.7071\tilde{y} + 0.2566) \nonumber \\
& & - 0.7071(0.7071\tilde{y} - 0.7071\tilde{x} + 0.2566)(p + r + 1.414\tilde{y} - 0.1623) \nonumber \\
& & - 1.414(0.7071\tilde{x} + 0.7071\tilde{y} + 0.2566)(0.7071\tilde{y} - 0.7071\tilde{x} + 0.2566)\nonumber \\
& & - 0.7071(r + 0.1623)(0.7071\tilde{y} - 0.7071\tilde{x} + 0.2566) \nonumber \\
& & +1.414(p + r + 1.414\tilde{y} - 0.1623)^2 
\end{eqnarray}

Next, we use a quadratic approximation for the center manifold of the above system \citesi{Carr} i.e. we assume $\tilde{y}=h(\tilde{x}) = \frac{1}{2} w \tilde{x}^2$. We can find $w$ by comparing two expressions obtained for $\frac{d\tilde{y}}{dt}$; the first is obtained by using  $\frac{d\tilde{y}}{dt} = \frac{d\tilde{y}}{d\tilde{x}} \frac{d\tilde{x}}{dt}$ and then using the first equation in Eq.~\ref{eq:newSystem2} and the quadratic approximation for $\tilde{y}$; the second is obtained by direct substitution of the quadratic approximation into the second equation in Eq.~\ref{eq:newSystem2}. Doing this yields:

$$
		\tilde{y}=h(\tilde{x})=-0.7071\tilde{x}^2/(4p + 4r - 2.1620)
$$
Hence we obtain the following one dimensional system restricted to the one-dimensional center manifold:

\begin{eqnarray}
\label{eq:newSystem}
\frac{\partial \tilde{x}}{\partial t}&=&0.1814r - 0.1814p - \tilde{x}(1.5p + 1.5r) \nonumber \\
       & &+ 0.7071(1.5p + 0.2434)(p + r - 0.1623)\nonumber  \\
       & &- 0.7071(1.5r + 0.2434)(p + r - 0.1623) \nonumber \\
       & &- \tilde{x}^2( 0.7071(1.5p + 0.2434)/(4p + 4r - 2.1623) \\
       & &- 0.7071(1.5r + 0.2434)/(4p + 4r - 2.1623) \nonumber \\
       & &- 0.7071(p + 0.1623)/(8p + 8r - 4.3246) \nonumber \\
       & &+ 0.7071(r + 0.1623)/(8p + 8r - 4.3246)) \nonumber \\
       & &+ \tilde{x}^3/(4p + 4r - 2.1623) \nonumber
\end{eqnarray}

It is easy to check that the origin in this transformed system satisfies the necessary conditions for a cusp bifurcation. The origin of this transformed system corresponds to the point $p_A = p_B = p_c$ in our original system Eq.~\ref{SI_MF}. Thus, the system undergoes a cusp bifurcation at $p_A = p_B = p_c$ where $p_c = \sqrt{10}-3 \approx 0.1623$.

\section{Mapping out the bifurcation curves (first order transition lines)}
In order to map out the first-order transition line (bifurcation curve) we adopt a semi-analytical approach.
We assume $p_B = c p_A$ with $c < 1$ to obtain the lower bifurcation curve (symmetry of the system allows us to obtain the upper bifurcation curve, given the lower one). Using Eqs. \ref{SI_MF}, the fixed point condition becomes (for simplicity, we denote $p_A$ by $p$):
\begin{eqnarray*}
f(x,y,p) &\equiv& -x y + (1-x-y-(1+c)p)^2 + x (1-x-y-(1+c)p) \nonumber \\ &+& \frac{3}{2} p (1-x-y-(1+c)p) - cp x = 0 \nonumber \\
g(x,y,p) &\equiv& -x y + (1-x-y-(1+c)p)^2 + y(1-x-y-(1+c)p) \nonumber \\ &+& \frac{3}{2} cp (1-x-y-(1+c)p) - p y = 0 \nonumber \\
\label{bif_conds12}
\end{eqnarray*}

In addition, for a fold bifurcation, we also require that the stability matrix has an eigenvalue with zero real part. Since, the valid solutions in our case are always real, this is equivalent to requiring the determinant of the stability matrix to be zero. Thus the condition $|Q| = 0$ (with $Q$ given by Eq.~\ref{stabilitymat}) along with Eqs. ~\ref{bif_conds12} enable us to determine for a given $c$, the location $(p_A,cp_A)$ at which the bifurcation occurs. By numerically solving these equations for different values of $c$, $0<c\le 1$ at intervals of $0.1$, we obtain the lower bifurcation curve shown in Fig. 2 of the main text.

\section{Optimal fluctuational paths, the eikonal approximation and switching times between co-existing stable states}

The master equation for our system takes the general form:
\[ \frac{\partial P({\bf X},t)}{\partial t} = \sum_{{\bf r}} \bigg[ W({\bf X}-{\bf r},{\bf r}) P({\bf X}-{\bf r},t) - W({\bf X},{\bf r}) P({\bf X},t) \bigg] \]
where ${\bf X} = [N_A \; N_B]^T$ denotes the (macro) state of the system as vector whose elements are the numbers of uncommitted nodes in state $A$ and $B$ respectively, $W({\bf X},{\bf r})$ is the probability of the transition from ${\bf X}$ to ${\bf X}+ {\bf r}$,  and ${\bf r}$ runs over the allowed set of displacement vectors in the space of macro-states. For our system, ${\bf r}$ runs over $[1\; 0]^T , [0 \; 1]^T, [2 \; 0]^T, [0 \; 2]^T, [-1 \; 0]^T, [0 \; -1]^T$. 
The deterministic equations can be derived from this master equation and yield:
\[
\frac{d {\bf X}_{\rm det}}{dt} = \sum_{\bf r} {\bf r} W( {\bf X}_{\rm det}, {\bf r})
\]

The Wentzell-Friedlin theory \citesi{Wentzell1984,Maier1992} assumes that for any path $[{\bf X}]$ in configuration space:
\[
\mathcal{P}([{\bf X}]) \sim \exp(-\mathcal{S}([{\bf X}]))
\]
with $\mathcal{S}([{\bf X}^*]) = 0$ for the deterministic path $[ {\bf X^*}]$.
It follows that the dominant contribution to the probability of a fluctuation that brings the system to state ${\bf X}$ starting from a stable state ${\bf X}_m$ can be written as:
\begin{equation}
\mathcal{P}({\bf X}| {\bf X}_m, t = 0) = \exp(-S({\bf X}))
\label{eikonal}
\end{equation}
where 
\begin{equation}
S({\bf X}) = min_{[\bf{X}]:{\bf X}_m \rightarrow {\bf X}} \mathcal{S}([{\bf x}])
\label{action1}
\end{equation}
where the minimization is over all paths $[{\bf X}]$ starting at ${\bf X}_m$ and ending at ${\bf X}$. For ${\bf X}$ far away from the steady state, the probability of occupation $P(\bf{X})$ is equivalent to logarithmic accuracy to the probability of the most likely fluctuation, $\mathcal{P}({\bf X}| {\bf X}_m, t = 0)$ that brings the system to ${\bf X}$. The assumption of the form given by Eq.~\ref{eikonal} for the occupation probability is known as the eikonal approximation.

Using a smoothness assumption for $W({\bf X},{\bf r})$, and since the changes in numbers of $A$ and $B$ nodes are $O(1)$, we can neglect the difference between $W({\bf X}-{\bf r},{\bf r})$ and $W({\bf X},{\bf r})$. With this approximation, the eikonal form for the occupation probabilities in the master equation yields the following equation for $S({\bf X})$ \citesi{Dykman1994}:
\begin{equation}
H\bigg({\bf x}, \frac{\partial s}{\partial {\bf x}}\bigg) = 0
\label{HJ}
\end{equation}
where 
\begin{equation}
H({\bf x},{\bf p}) = \sum_{\bf r}  w({\bf x},{\bf r}) (\exp({\bf r}{\bf p})-1)
\label{Hamiltonian}
\end{equation}
 and
 \[
 {\bf x} = {\bf X}/N, \; \;  w({\bf x},{\bf r}) = W({\bf X},{\bf r})/N, \; \; s({\bf x}) = S({\bf X})/N.
 \]
Eq.~\ref{HJ}, is analogous to a Hamilton-Jacobi equation for the action of a system with Hamiltonian given by Eq.~\ref{Hamiltonian}.
The corresponding Hamilton equations of motion for components of position ${\bf x}$ and momentum ${\bf p}$ are:
\begin{equation}
\dot{x_i} = \frac{\partial H}{\partial p_i} \; \;  \;  \dot{p_i} = -\frac{\partial H}{x_i}
\label{eqsofmot}
\end{equation}
with $s({\bf x})$ playing the role of the classical action:
\[
s([{\bf x}]) = \int_{[{\bf x}]} L({\bf x},{\bf p}) d{\bf x} = \int_{[{\bf x}]}  {\bf p} \dot{{\bf x}} d{\bf x}
\]
where $[{\bf x}]$ denotes a particular path obeying the equations of motion (Eqs.~\ref{eqsofmot}).

Following this Hamiltonian formulation to characterize the fluctuational paths of the system, our goal is to find the path with minimum action that reaches the separatrix in phase space of the deterministic motion, starting from the vicinity of the stable state under consideration \citesi{Maier1992,Dykman1994}. Arguments in \citesi{Dykman1979} show that the fluctuational path reaching the separatrix with the minimal value of the action, is the path that passes through the saddle point. This is the {\it optimal escape path}, i.e., the path whose probability of occurrence dominates the probability of escape and we denote it by $[{\bf x}_{\rm opt}]$.
This path can be found by integrating the equations of motion Eq.~\ref{eqsofmot}, and finding the required path that starts from the vicinity ${\bf x}_m$ to the saddle point ${\bf x}_{\rm saddle}$. Thus following Eqs.~\ref{eikonal},~\ref{action1} we have for the probability of escape from the current stable point in which the system is trapped:
\begin{equation}
P_{\rm escape} = P({\bf x}_{\rm saddle}) \sim \exp[-N s({\bf x}_{\rm saddle})]
\label{escapeprob}
\end{equation}
where
\[
s({\bf x}_{\rm saddle}) = \int_{[{\bf x}_{\rm opt}]} L({\bf x},{\bf p}) d{\bf x}
\]
and the transition time (or time to escape from the steady state) follows:
\begin{equation}
T_{\rm switching} \sim \exp[ N s({\bf x}_{\rm saddle})]
\label{switching}
\end{equation}

In practice we start from some point ${\bf x}$ in the vicinity of the stable state, and to obtain the corresponding momenta ${\bf p}$ and action $s({\bf x})$, we employ a Gaussian approximation \citesi{Dykman1994}:
\[
S({\bf x}) = \sum Z_{ij} (x_i - x_i^m) (x_j - x_j^m)
\]
where $Z$ satisfies an algebraic Ricatti equation:
\[
{\bf Q}{\bf Z}^{-1} + {\bf Z}^{-1}{\bf Q}^T + {\bf K} = 0
\]
where $Q$ is the linear stability matrix (Eq.~\ref{stabilitymat}) evaluated at ${\bf x}_m$ and  $K_{ij} = \sum_{\bf r} w({\bf x}_m,{\bf r}) r_i r_j$.
Solving this Ricatti equation yields $Z$ which in turn yields $S({\bf x})$ and $p({\bf x})$.

In order to find the optimal fluctuational path of escape from a given steady state, we numerically generate fluctuational paths from various points close to the steady state (we explore points at intervals of $10^{-5}$ along the $x_1$ dimension and $10^{-2}$ along the $x_2$ dimension around the steady state) and find one that passes close enough (no greater than a distance of $10^{-5}$) to the saddle point. The equations of motion, Eqs.~\ref{eqsofmot}, are integrated using a trapezoidal rule to generate these paths starting with initial conditions obtained using the Gaussian approximation described above and subsequent numerical solution of the Ricatti equation (we use a Matlab Ricatti equation solver for the latter). The scaling behavior of switching times obtained using this approach for various committed fraction values as a function of distance from second-order transition (or cusp) point are shown in Fig. 5 of the main text.

%
%

\end{document}